\documentclass[10pt, journal]{IEEEtran}
\IEEEoverridecommandlockouts

\newcommand{\Design}{P2LSG}

\usepackage{flushend}

\usepackage{color}
\usepackage{xcolor}

\usepackage{amssymb}

\definecolor{bronze}{rgb}{0.8, 0.5, 0.2}

\usepackage{tikz}

\usepackage{amsmath}
\usepackage{amsfonts}
\usepackage{amssymb}
\usepackage{mathtools}

\usepackage[hidelinks,colorlinks=true,linkcolor=blue,citecolor=blue]{hyperref}

\usepackage{xcolor}
\usepackage{circledsteps}
\newcommand\myCircled[2][]{\ifmmode
\Circled[fill color=black,inner color=white,#1]{\mathsf{#2}}
\else
\Circled[fill color=black,inner color=white,#1]{\sffamily#2}
\fi
}

  \renewenvironment{thebibliography}[1]{%
    \begin{oldthebibliography}{#1}%
      \setlength{\parskip}{-0.0ex}%
      \setlength{\itemsep}{-0.0ex}%
  }%
  {%
    \end{oldthebibliography}%
  }

\usepackage{cite}
\usepackage{amsmath,amssymb,amsfonts}
\usepackage{algorithmic}
\usepackage{graphicx}
\usepackage{textcomp}
\usepackage{xcolor}

\usepackage[left=0.59in, right=0.59in, top=0.60in, bottom=0.60in]{geometry}

\def\BibTeX{{\rm B\kern-.05em{\sc i\kern-.025em b}\kern-.08em
    T\kern-.1667em\lower.7ex\hbox{E}\kern-.125emX}}
 
\newcommand{\ignore}[1]{ }

\newcommand*\circled[1]{\tikz[baseline=(char.base)]{
            \node[shape=circle,draw,inner sep=1.0pt] (char) {#1};}}

\usepackage{multirow}
\usepackage{vcell}

\usepackage{comment}

\usepackage{amsmath}
\usepackage{mathtools}

\usepackage{ragged2e}

\usepackage{pifont}

\usepackage{multicol}

\usepackage{enumitem}

\usepackage[hidelinks]{hyperref}

\usepackage[normalem]{ulem}

\usepackage{makecell}

\usepackage{algorithm}
\usepackage{algorithmic}

\usepackage{colortbl}
\usepackage{hhline}

\usepackage{marvosym}

\usepackage{nccmath}

\newcommand{\blue}{}

\renewcommand{\blue}{\textcolor{blue}}

\usepackage[prependcaption, textsize=footnotesize]{todonotes}

\newcommand*\titleheader[1]{\gdef\@titleheader{#1}}

\usepackage{tikz,xcolor}

\definecolor{lime}{HTML}{A6CE39}
\DeclareRobustCommand{\orcidicon}{%
	\begin{tikzpicture}
	\draw[lime, fill=lime] (0,0) 
	circle [radius=0.16] 
	node[white] {{\fontfamily{qag}\selectfont \tiny ID}};
	\draw[white, fill=white] (-0.0625,0.095) 
	circle [radius=0.007];
	\end{tikzpicture}
	\hspace{-2mm}
}

\foreach \x in {A, ..., Z}{%
	\expandafter\xdef\csname orcid\x\endcsname{\noexpand\href{https://orcid.org/\csname orcidauthor\x\endcsname}{\noexpand\orcidicon}}
}

\usepackage{courier}

\begin{document}

\title{P2LSG: Powers-of-2 Low-Discrepancy\\ Sequence Generator for Stochastic Computing
\vspace{-4mm}
}

\author{
        Mehran~Shoushtari~Moghadam\orcidA{},
        Sercan~Aygun\orcidB{},
        Mohsen~Riahi~Alam\orcidC{},
        M.~Hassan~Najafi\orcidD{}

\vspace{-20pt}

\thanks{The authors are with the School of Computing and Informatics, Center for Advanced Computer Studies, University of Louisiana at Lafayette, Lafayette, LA 70503, USA (E-mail: \{mehran.shoushtari-moghadam1, \linebreak sercan.aygun, mohsen.riahi-alam, najafi\}@louisiana.edu). This work was supported in part by National Science Foundation (NSF) grant \#2019511, the Louisiana Board of Regents Support Fund \#LEQSF(2020-23)-RD-A-26, and generous gifts from Cisco, Xilinx, and NVIDIA.}

}

\markboth{\small{\ \ \ \ \ \ \ \ \ \ \ \ \ \ \ \ \ \ \   T\lowercase{his} \lowercase{work was accepted to the 29$^{th}$} A\lowercase{sia and} S\lowercase{outh} P\lowercase{acific} D\lowercase{esign} A\lowercase{utomation} C\lowercase{onference} (ASP-DAC) 2024}}%
{\MakeLowercase{\textit{}}}

\titleheader{2016 IEEE 24th International Requirements Engineering Conference}

\maketitle

\begin{abstract}
Stochastic Computing (SC) is an unconventional computing paradigm processing data in the form of random bit-streams.
The accuracy and energy efficiency of SC systems highly depend on the stochastic number generator (SNG) unit that converts the data from conventional binary to stochastic bit-streams. Recent work has shown significant improvement in the efficiency of SC systems by employing low-discrepancy (LD) sequences such as \textbf{\texttt{Sobol}} and \textbf{\texttt{Halton}} sequences in the SNG unit.  
Still, the usage of many well-known random sequences for SC remains unexplored. 
This work studies some new random sequences for potential application in SC. 
Our design space exploration proposes 
a promising random number generator for accurate and energy-efficient SC. 
{We propose \textbf{\texttt{P2LSG}}, a low-cost and energy-efficient 
Low-discrepancy Sequence Generator derived from \textit{Powers-of-2} VDC (Van der Corput) sequences.}
We evaluate the performance of {our novel} bit-stream {generator} for two SC image and video processing case studies: image scaling and scene merging. For the scene merging task, we propose a novel SC design for the first time. Our experimental results show higher accuracy and lower hardware cost and energy consumption compared to the state-of-the-art.  
\end{abstract}


\vspace{-0.5em}

\section{Introduction}
\label{intro}

\IEEEPARstart{S}{tochastic} computing (SC)~\cite{Gaines1969} is a re-emerging computing paradigm offering low-cost and noise-tolerant hardware designs. 
In contrast to traditional binary computing, which operates on positional binary radix numbers, SC designs process uniform bit-streams of `\MVZero's and `\MVOne's with no significant digits. While the paradigm was known for approximate computations for years, recent works showed deterministic and completely accurate computation using SC circuits~\cite{Najafi_TVLSI_2019, ExactStoch21}. Encoding data from {traditional} binary to stochastic bit-streams is an important step in any SC system. The data are encoded by the probability of observing a `\MVOne' in the bit-stream. For example, a bit-stream with 25\% `\MVOne' represents the data value of 0.25. The accuracy of the computations and the energy efficiency of the SC designs highly depend on this encoding step, particularly on the distribution of `\MVOne's and `\MVZero's in the bit-streams.
A stochastic number generator (SNG), which encodes a data value in binary format to a stochastic bit-stream, consists of a random number generator (RNG) and a binary comparator. Fig.~\ref{sng_fig} shows the structure of an SNG commonly used in SC systems. At any cycle, the output of comparing the input data with the random number from the RNG unit produces one bit of the bit-stream.

The choice of the RNG unit directly affects the distribution of the bits in stochastic bit-streams.
While traditionally \textit{pseudo-random} sequences, generated by linear-feedback shift registers (LFSRs), were used in the SNG unit, the state-of-the-art (SOTA) studies employ 
\textit{quasi-random} sequences, such as \textbf{\texttt{Sobol}} (\textbf{\texttt{S}})~\cite{7927069,Najafi_TVLSI_2019} and \textbf{\texttt{Halton}} (\textbf{\texttt{HL}})~\cite{9406123, 6800290} sequences, for high-quality generation of stochastic bit-streams. 
These sequences remove an important source of error in SC, 
the random fluctuation error~\cite{8244436} in generating bit-streams by producing \textit{Low-Discrepancy} (LD) bit-streams.
LD bit-streams quickly converge to the target value, reducing the length of bit-streams and, consequently, the latency of stochastic computations. This latency reduction 
directly translates to  savings in energy consumption (i.e., \linebreak power $\times$ latency), a critical metric in the hardware efficiency of SC systems.

A challenge with the SOTA 
SNGs using sequences such as \textbf{\texttt{Sobol}} and \textbf{\texttt{Halton}} is their relatively high hardware cost that affects 
the 
achievable energy savings when using these sequences. This study extends the SOTA 
random sequences for the high-quality encoding of data in SC. 
We analyze some 
high-quality random sequences 
for possible improvement in the performance and hardware efficiency of the SNG unit. 
For the first time, to the best of our knowledge, we explore \textbf{\texttt{Weyl}} (\textbf{\texttt{W}}) \cite{W}, \textbf{\texttt{R2}} (\textbf{\texttt{R}}) \cite{R}, 
\textbf{\texttt{Latin Hypercube}} (\textbf{\texttt{L}}) \cite{L}, 
\textbf{\texttt{Faure}} (\textbf{\texttt{F}}) \cite{F}, 
\textbf{\texttt{Hemmersly}} (\textbf{\texttt{HM}}) \cite{H}, 
\textbf{\texttt{Niederreiter}} (\textbf{\texttt{N}}) \cite{N,N2},
\textbf{\texttt{Van der Corput}} (\textbf{\texttt{VDC}})~\cite{V},
and \textbf{\texttt{Poisson Disk}} (\textbf{\texttt{P}}) \cite{P} sequences in the context of SC 
as promising alternatives to prior costly LD sequences. 
The primary contributions of this work are summarized as follows:

\begin{figure}[t]
  \centering
  \includegraphics[width=180pt]{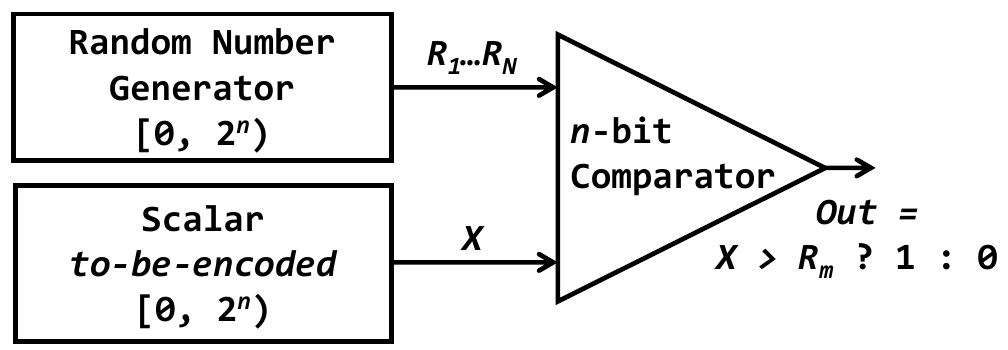}
\vspace{-1em}
  \caption{The architecture of a stochastic number generator (SNG).}
  \label{sng_fig}
\end{figure}

\noindent \textbf{\circled{1}} We employ some high-quality random sequences 
for the first time in SC literature. We evaluate the performance and 
accuracy of stochastic computations when using these sequences. 

\noindent \textbf{\circled{2}} We propose \textbf{\texttt{P2LSG}} (\textbf{\texttt{P}}owers-of-\textbf{\texttt{2}} \textbf{\texttt{L}}ow-Discrepancy
\textbf{\texttt{S}}equence \textbf{\texttt{G}}enerator), 
a lightweight LD Sequence Generator derived from \textit{Powers-of-2} \textbf{\texttt{VDC}} sequences 
and evaluate its hardware cost compared to the SOTA. 

\noindent \textbf{\circled{3}} For the first time, we introduce a novel SC video processing design for scene merging. 

\noindent \textbf{\circled{4}} 
Experimental results on two SC image and video processing case studies show significant savings in area, energy, and latency with \textbf{\texttt{P2LSG}} compared to the SOTA non-parallel and parallel \textbf{\texttt{Sobol}}-based designs 
while maintaining comparative accuracy.

The rest of the paper is structured as follows.
Section~\ref{background_and_motivation} provides the necessary background on random sequences and SC. 
Section~\ref{design_space} explores different random sequences 
in the context of SC. Section~\ref{new_seq_des} reveals the performance of 
\textbf{\texttt{P2LSG}}
along with a new hardware design. 
Section~\ref{sc_image} investigates the proposed generator for image scaling (interpolation) and scene merging. 
Finally, Section~\ref{conclusions} concludes the paper, summarizing the key findings and contributions.

\section{Background}
\label{background_and_motivation}

\subsection{Random Sequences}
{Random sequences are widely 
used in various research domains, particularly in emerging computing technologies \cite{aygun2023learning}.
Some sequences are binary-valued with logic-\MVOne s and logic-\MVZero s \cite{mitra2008pseudo}. These sequences possess the orthogonality property; that is, 
different random sequences are (approximately) uncorrelated. Some sequences, on the other hand, are non-binary-valued (having fixed- or floating-point numbers). All non-binary-valued sequences listed in 
Section I (\textbf{\texttt{S}}, \textbf{\texttt{HL}}, \textbf{\texttt{W}}, \textbf{\texttt{R}}, \textbf{\texttt{L}}, \textbf{\texttt{F}}, \textbf{\texttt{HM}}, \textbf{\texttt{N}}, \textbf{\texttt{VDC}}, \textbf{\texttt{P}}) 
have LD properties. The \textit{discrepancy} term in LD refers to how much the sequence points deviate from uniformity~\cite{Sobol_TVLSI_2018}. The \textit{recurrence} property (i.e., the constructibility of further-indexed sequences from the previous-indexed ones) in LD sequences is beneficial for cross-correlation, which is} 
advantageous for SC 
systems that require uncorrelated bit-streams~\cite{Alaghi_SCC1}.

The \textbf{\texttt{Weyl}} sequence belongs to the class of additive recurrence sequences, characterized by their generation through the iteration of multiples of an irrational number modulo $1$. Specifically, by considering $\alpha \in \mathbb{R}$ as an irrational number and {\small $x_i \in \{0,\alpha,2\alpha,...,k\alpha\}$}, the sequence  {\small $x_i-\lfloor x_i \rfloor$} ($x_i$ modulo $1$) produces an equidistributed sequence within the interval {\small $(0,1)$}. Another example of an additive recurrence sequence is the \textbf{\texttt{R}} sequence, which is based on the \textit{Plastic Constant} (the unique real solution of the cubic equation) \cite{PlasticNC, R}. The \textbf{\texttt{Latin Hypercube}} sequences involve partitioning the sampling space into equally sized intervals and randomly selecting a point within each interval~\cite{lin2022latin}. 

The \textbf{\texttt{VDC}} sequence serves as the foundation for many LD sequences. It is constructed by reversing the digits of the numbers in a 
specific base, 
representing each integer value as a fraction within the {\small$[0,1)$} interval.
{
A \textbf{\texttt{VDC}} sequence in base-\texttt{B} is notated with \textbf{\texttt{VDC-B}}. 
As an example, the decimal value $11$ in base-3 is represented by 
{\small $(102)_3$}. The corresponding value for the 
\textbf{\texttt{VDC-3}} is {\small $2\times3^{-1}+0\times3^{-2}+1\times3^{-3} = \frac{19}{27}$}. Similarly, the \textbf{\texttt{Faure}}, \textbf{\texttt{Hammersley}}, and \textbf{\texttt{Halton}} sequences are derived from the \textbf{\texttt{VDC}} concept using prime or co-prime numbers. 
To generate the \textbf{\texttt{Faure}} sequence in $q$-dimensions, the smallest prime number $p$ is selected such that {\small $p \ge q$}. The first dimension of the \textbf{\texttt{Faure}} sequence corresponds to the \textbf{\texttt{VDC-p}} sequence, 
 while the remaining dimensions involve permutations of the first dimension. The $q$-dimensional \textbf{\texttt{Halton}} sequence is generated by utilizing the \textbf{\texttt{VDC}} sequence with different prime bases starting from the 
first 
to the $q$-th prime number.  
{A limitation of the \textbf{\texttt{Halton}} sequence is in utilizing prime number bases, which increases the complexity of the sequence 
generation~\cite{Krykova2004}.}

The \textbf{\texttt{Hammersley}} sequence shares some similarities with the \textbf{\texttt{Halton}} sequence. 
For the sake of fair comparison with the \textbf{\texttt{Halton}} sequence we adopt different bases for the \textbf{\texttt{Hammersley}} sequence in this work.
The first 
\textbf{\texttt{Sobol}} sequence is the same as the
\textbf{\texttt{VDC-2}} sequence. The other \textbf{\texttt{Sobol}} sequence 
are generated through permutations of some sets of direction vectors~\cite{sobol_sim}. The \textbf{\texttt{Niederreiter}} sequence is another variant of the \textbf{\texttt{VDC}} sequence relying on the powers of some prime numbers. This sequence features 
irreducible and primitive polynomials that ensure LD 
and uniformity over the sample space \cite{finite}. Finally, the \textbf{\texttt{Poisson Disk}} sequence 
generates 
evenly distributed numbers 
with minimal distance between them.

\subsection{Stochastic Computing (SC)}
SC 
has gained attention 
recently
due to its intriguing advantages, such as robustness to noise, high parallelism, 
and low design cost. Complex arithmetic operations are realized with simple logic gates in SC. Significant savings in the implementation costs are achieved for different applications, from image processing~\cite{SC_Image_2014} to sorting~\cite{najafiSortingUnary} and machine learning~\cite{ZheJiAoHEIF, AygunDissertation}, to name a few. 
Data conversion is an essential step in SC systems. 
Input numbers 
must be converted to random 
bit-streams, where each bit 
has equal significance. 
SC supports real data in the unit interval, i.e., [0, 1]. 
A common coding format is unipolar encoding (UPE).
In UPE, the probability of observing a `\MVOne' in the  bit-stream $X$, i.e., {\small $P(X=1)$}, equals the input value or $x$. 
The common 
method 
for generating a bit-stream of size $N$ is to compare the input number with $N$ random numbers {\small ($R_{1}...R_{N}$)}. This is usually done serially in $N$ clock cycles. 
A logic-\MVOne~is produced at the output 
if the input 
value is greater than the random number; A logic-\MVZero~is produced otherwise. 
The distribution 
of logic-\MVOne s in the produced bit-stream depends on the sequence of random numbers. 
When dealing with the signed values ($x$ is in the range {\small $-1 \leq x \leq 1$}),  
a bipolar encoding (BPE) is used, in which 
the probability that each bit in the bit-stream is `\MVOne' is {\small $P(X = 1)$ = $\frac{(x + 1)}{2}$}.

\begin{table*}
\centering
\caption{MAE ($\%$) comparison of using different random sequences when multiplying two 8-bit precision\\ stochastic bit-streams 
with different bit-stream lengths. 
}
\vspace{-0.8em}
\begin{tabular}{|c|c|c|c|c|c|c|c|c|c|c|c|} 
\hline
\textbf{Random Sequence} & \textbf{2\textsuperscript{6}} & \textbf{2\textsuperscript{7}} & \textbf{2\textsuperscript{8}} & \textbf{2\textsuperscript{9}} & \textbf{2\textsuperscript{10}} & \textbf{2\textsuperscript{11}} & \textbf{2\textsuperscript{12}} & \textbf{2\textsuperscript{13}} & \textbf{2\textsuperscript{14}} & \textbf{2\textsuperscript{15}} & \textbf{2\textsuperscript{16}} \\ 
\hline
\textbf{\texttt{Sobol}} & 0.92 & 0.45 & 0.19 & 0.092 & 0.041 & 0.019 & 0.009 & 0.0035 & 0.0013 & 0.0003 & 0.0000 \\ 
\hline
\textbf{\texttt{R2}} & 1.14 & 1.07 & 0.48 & 0.220 & 0.130 & 0.055 & 0.037 & 0.0164 & 0.0099 & 0.0078 & 0.0024 \\ 
\hline
\textbf{\texttt{Weyl}} & 1.46 & 1.40 & 0.83 & 0.530 & 0.400 & 0.220 & 0.190 & 0.1800 & 0.1300 & 0.0090 & 0.0065 \\ 
\hline
\textbf{\texttt{Latin Hypercube}} & 3.19 & 0.93 & 0.85 & 0.380 & 0.390 & 0.250 & 0.240 & 0.1200 & 0.0795 & 0.0508 & 0.0424 \\ 
\hline
\textbf{\texttt{Faure}} & 2.60 & 1.40 & 0.88 & 0.480 & 0.210 & 0.110 & 0.077 & 0.0360 & 0.0136 & 0.0113 & 0.0040 \\ 
\hline
\textbf{\texttt{Halton}} & 3.31 & 1.42 & 1.14 & 0.780 & 0.380 & 0.150 & 0.093 & 0.0570 & 0.0330 & 0.0150 & 0.0083 \\ 
\hline
\textbf{\texttt{Hammersley}} & 1.31 & 0.85 & 0.37 & 0.200 & 0.120 & 0.061 & 0.030 & 0.0170 & 0.0098 & 0.0043 & 0.0019 \\ 
\hline
\textbf{\texttt{Niederreiter}} & 0.95 & 0.51 & 0.34 & 0.130 & 0.072 & 0.032 & 0.019 & 0.0067 & 0.0039 & 0.0015 & 0.0011 \\ 
\hline
\textbf{\texttt{Poisson Disk}} & 4.16 & 2.61 & 1.06 & 0.960 & 0.710 & 0.340 & 0.440 & 0.3800 & 0.3700 & 0.6000 & 0.5200 \\
\hline
\textbf{\texttt{P2LSG}} & 1.76 & 0.88 & 0.39 & 0.170 & 0.073 & \textbf{0.030} & \textbf{0.012} & \textbf{0.0045} & \textbf{0.0015} & \textbf{0.0003} & \textbf{0.0000} \\ 
\hline

\end{tabular}
\label{Table2_mul}
\vspace{-0.7em}
\justify{\scriptsize{
The evaluation was conducted for 8-bit precision input data, 
and accurate multiplication results were obtained after $2^{2\times 8}$ operation cycles. The assessment was performed for operation cycles ranging from $2^6$ to $2^{16}$, corresponding to 
bit-stream lengths varying from $2^6$ to $2^{16}$.
In the experiment, the first two \textbf{\texttt{Sobol}} sequences were extracted from \texttt{MATLAB}. The $\pi$ and \textit{Silver Ratio} numbers were employed to generate the \textbf{\texttt{Weyl}} sequence. The \textbf{\texttt{Faure}} sequence was generated using \textbf{\texttt{VDC-7}}. 
The \textbf{\texttt{Halton}} sequence was utilized with \textbf{\texttt{VDC-11}} and \textbf{\texttt{VDC-13}}. The \textbf{\texttt{Hammersley}} sequence was constructed using \textbf{\texttt{VDC-2}} and \textbf{\texttt{VDC-3}}. For the \textbf{\texttt{P2LSG}} sequence, at each operation cycle $i$, the \textbf{\texttt{VDC-2}} and \textbf{\texttt{VDC-$2^i$}} were employed. \blue{\textbf{Our framework can be found in~\cite{our_github}}}.
}}
\end{table*}

\begin{figure*}[t]
\label{fig3}
  \centering
  \includegraphics[width=450pt]{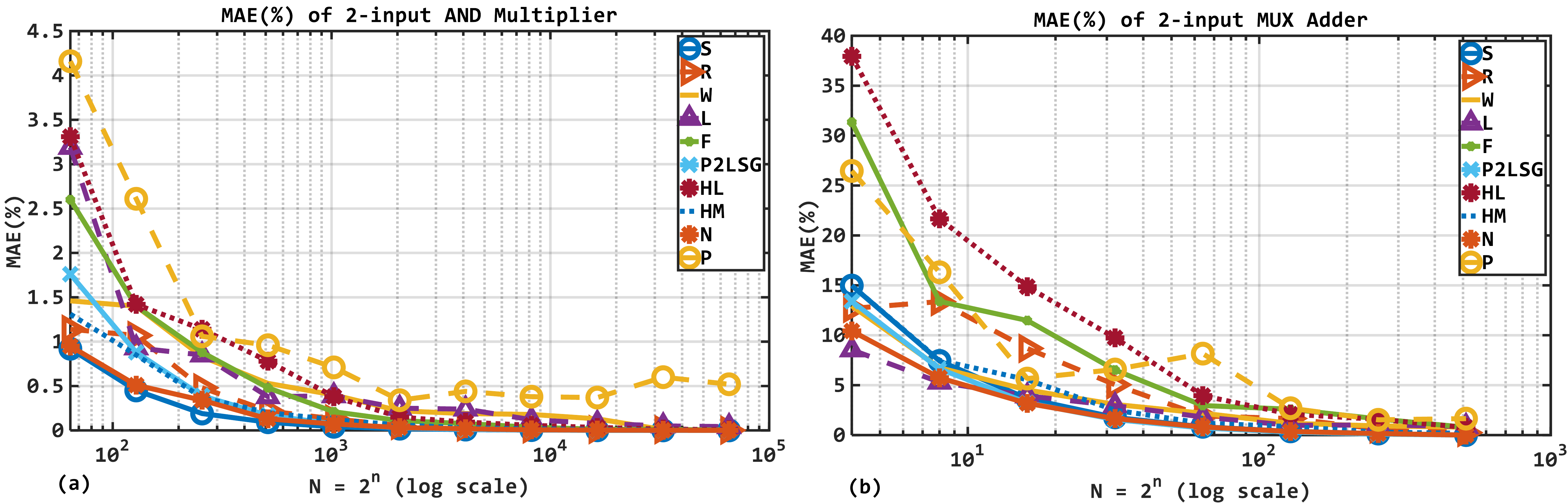}
\vspace{-0.5em}
  \caption{MAE ($\%$) of SC operation on two 8-bit precision input  (a) SC Multiplication and (b) SC Scaled Addition. \textbf{\texttt{VDC}}-related sequences demonstrate favorable convergence rates. 
  For approximate results, the \textbf{\texttt{S}}, \textbf{\texttt{N}}, \textbf{\texttt{HM}}, and \textbf{\texttt{P2LSG}} sequences emerge as the top performers.
  }
  \label{benchmark_and_mux}
\vspace{-0.9em}
\end{figure*}

\begin{table}
\centering
\small
\caption{MAE ($\%$) comparison of the two 8-bit precision input bit-streams for SC Scaled Addition with different bit-stream lengths. 
}
\vspace{-0.8em}
\setlength{\tabcolsep}{1.7pt}
\resizebox{3.5in}{!}{
\begin{tabular}{|c|c|c|c|c|c|c|c|c|} 
\hline
\textbf{Seq.} & \textbf{2\textsuperscript{2}} & \textbf{2\textsuperscript{3}} & \textbf{2\textsuperscript{4}} & \textbf{2\textsuperscript{5}} & \textbf{2\textsuperscript{6}} & \textbf{2\textsuperscript{7}} & \textbf{2\textsuperscript{8}} & \textbf{2\textsuperscript{9}} \\ 
\hline
\textbf{\texttt{Sobol}} & 14.98 & 7.43 & 3.66 & 1.77 & 0.83 & 0.37 & 0.15 & 0.00 \\ 
\hline
\textbf{\texttt{R2}} & 12.69 & 13.37 & 8.68 & 5.04 & 1.91 & 1.66 & 0.60 & 0.25 \\ 
\hline
\textbf{\texttt{Weyl}} & 12.94 & 6.74 & 4.52 & 3.07 & 2.19 & 1.17 & 0.92 & 0.79 \\ 
\hline
\textbf{\texttt{Latin Hypercube}} & 8.54 & 5.30 & 3.92 & 2.95 & 1.81 & 0.97 & 1.02 & 0.86 \\ 
\hline
\textbf{\texttt{Faure}} & 31.35 & 13.39 & 11.48 & 6.52 & 2.96 & 2.59 & 1.62 & 0.78 \\ 
\hline
\textbf{\texttt{Halton}} & 37.94 & 21.66 & 14.86 & 9.74 & 3.95 & 2.09 & 1.58 & 0.75 \\ 
\hline
\textbf{\texttt{Hammersley}} & 14.98 & 7.49 & 5.55 & 2.49 & 1.19 & 0.85 & 0.29 & 0.18 \\ 
\hline
\textbf{\texttt{Niederreiter}} & 10.43 & 5.74 & 3.18 & 1.65 & 0.81 & 0.36 & 0.15 & 0.00 \\ 
\hline
\textbf{\texttt{Poisson}} & 26.46 & 16.31 & 5.66 & 6.57 & 8.16 & 2.69 & 1.53 & 1.65 \\
\hline
\textbf{\texttt{P2LSG}} & 13.40 & 6.63 & 3.24 & \textbf{1.55} & \textbf{0.71} & \textbf{0.29} & \textbf{0.097} & \textbf{0.00} \\ 
\hline

\end{tabular}
}
\label{Table3_add}
\vspace{-0.6em}
\justify{\scriptsize{

For 8-bit precision input data, accurate scaled addition results were obtained after $2^{8+1}$ operation cycles. The evaluation was conducted for operation cycles ranging from $2^2$ to $2^{9}$.
In the experiment, the first two \textbf{\texttt{Sobol}} sequences were extracted from \texttt{MATLAB}. The $\pi$ and \textit{Silver Ratio} numbers were employed to generate the \textbf{\texttt{Weyl}} sequence. The \textbf{\texttt{Faure}} sequence was generated using \textbf{\texttt{VDC-7}}. 
The \textbf{\texttt{Halton}} sequence was utilized with \textbf{\texttt{VDC-11}} and \textbf{\texttt{VDC-13}}. The \textbf{\texttt{Hammersley}} sequence was constructed using \textbf{\texttt{VDC-2}} and \textbf{\texttt{VDC-3}}. For 
\textbf{\texttt{P2LSG}}, 
at each operation cycle $i$, the \textbf{\texttt{VDC-2}} and \textbf{\texttt{VDC-$2^i$}} were employed.
\vspace{-5pt}
}}
\end{table}

SC operations consist of simple bit-wise logic operations. 
Multiplication of bit-streams in UPE is achieved by 
bit-wise \texttt{AND}~\cite{Alaghi_SCC1}, and in BPE by bit-wise \texttt{XNOR} operation~\cite{7965993}.
For accurate multiplication, the input bit-streams must be \textit{uncorrelated} with each other. Performing bit-wise \texttt{AND} on \textit{correlated} bit-streams 
with a maximum overlap in the position of `\MVOne's gives the \textit{minimum} of the input bit-streams~\cite{najafiSortingUnary}.
Scaled addition is realized in SC by 
a multiplexer (\texttt{MUX}) unit for both encodings~\cite{10.1145/3491213}. 
For scaled subtraction, a 
\texttt{MUX} with one inverter 
is utilized~\cite{SC_Image_2014}.
The main inputs of the \texttt{MUX} can be correlated, but they should be uncorrelated with the select input bit-stream.

\section{Design Space Exploration}
\label{design_space}
This section comprehensively 
examines the use of the random sequences discussed in Section~\ref{background_and_motivation} for SC. We first analyze these sequences for basic SC operations and then extend the evaluations to more complex case studies. 
The numbers provided by these sequences are used as the required random numbers ($R_{1}...R_{N}$) 
during bit-stream generation. 
Prior works 
 used \textbf{\texttt{Sobol}}~\cite{7927069} and \textbf{\texttt{Halton}}~\cite{6800290} sequences for LD bit-stream generation. 
In this study, for the first time we propose \textbf{\texttt{P2LSG}}, a 
new LD sequence generator based on the \textbf{\texttt{VDC}} \textit{Powers-of-2} bases (e.g., \textbf{\texttt{VDC-2}}, \textbf{\texttt{VDC-4}}, \textbf{\texttt{VDC-8}},..., \textbf{\texttt{VDC-2$^{m \in {\mathbb{Z}}^+}$}}) for cost-efficient LD bit-stream generation.
The proposed sequence generator is cost- (area and power) and energy-efficient for hardware implementation. 

\subsection{Benchmark-I: SC Multiplication}
\label{benchmark_1}
We first evaluate the performance of the selected sequences for 2-input SC multiplication. Two input values ({\small $X1$} and {\small $X2$}) are converted to bit-stream representation using random sequences, and the generated bit-streams are bit-wise \texttt{AND}ed to produce the output bit-stream. The resulting bit-stream is converted back to standard representation (by counting the number of `\MVOne's and dividing by the length of the bit-stream) and compared with the expected multiplication result to find the absolute error. 
Here, the expected value is {\small $P_{X1} \times P_{X2}$}. For accurate multiplication, the input bit-streams must be \textit{uncorrelated}. 
In the literature, 
\textit{Stochastic Cross-Correlation} ({\small $SCC$}) is used 
to quantify the correlation between bit-streams~\cite{Alaghi_SCC1}. 
In this metric, the correlation is calculated by using cumulative values denoted by $a$, $b$, $c$, and $d$, which depend on the counts of \MVOne\MVOne, \MVOne\MVZero, \MVZero\MVOne, or \MVZero\MVZero~pairs in the overlapping bits between  the two 
bit-streams: 
\begin{equation}
\label{SCC_equation}
    SCC = \begin{cases} \frac{ad-bc}{N \times min(a+b, a+c)-(a+b) \times(a+c)} & ,  \  if \ ad>bc \\ \frac{ad-bc}{(a+b) \times(a+c) - N \times max(a-d, 0)} & ,  \ else\end{cases}
\end{equation}

We exhaustively evaluated the multiplication accuracy for all cartesian combinations 
of {\small $X1$} and {\small $X2$} 
where the inputs are $8$-bit precision values in the {\small $[0,1)$} interval (i.e., {\small $\frac{0}{256}$}, {\small $\frac{1}{256}$},..., {\small $\frac{255}{256}$}). 
The bit-stream lengths vary from $2^6$ to $2^{16}$ with $2\times$ increments. Table~\ref{Table2_mul} and Fig.~\ref{benchmark_and_mux} (a) present
the Mean Absolute Error (MAE) of the multiplication results. 
We multiply the measured mean values by 100 and report them as percentages.
Two different sequences are selected for each case to satisfy the uncorrelation requirement ({\small $SCC=0$}). 
For the \textbf{\texttt{Sobol}} sequence, the first two \textbf{\texttt{Sobol}} sequences from the \texttt{MATLAB} built-in \textbf{\texttt{Sobol}} sequence generator are used. 
For the \textbf{\texttt{Faure}} sequence, two sequences are created using 
\textbf{\texttt{VDC-7}}.
The first dimension
is generated using base-7, while the second  
is obtained through a permutation of the first one. 
The \textbf{\texttt{Halton}} sequence involves two dimensions generated using base-11 (\textbf{\texttt{VDC-11}}) and base-13 (\textbf{\texttt{VDC-13}}) with 
\texttt{MATLAB} built-in \textbf{\texttt{Halton}} 
function. 
For the \textbf{\texttt{Hammersley}} sequence, we use the \textbf{\texttt{VDC-2}} and \textbf{\texttt{VDC-3}} 
sequences. 
The \textbf{\texttt{Latin Hypercube}} sequence was also generated using its \texttt{MATLAB} built-in function. 
For the \textbf{\texttt{Weyl}} sequence, $\pi$ and the \textit{Silver Ratio} (i.e. {\small $\sqrt{2}-1$}) were chosen as the irrational numbers. The first dimension of our \textbf{\texttt{P2LSG}} sequence is 
\textbf{\texttt{VDC-2}}, while \textbf{\texttt{VDC-N}} is selected for the other dimensions depending on  
the bit-stream length ($N$). 
As we move from left to right in Table~\ref{Table2_mul}, the length of the bit-streams increases, and as expected, the accuracy 
improves (MAE decreases). 
Notably, the \textbf{\texttt{VDC}}-related sequences exhibit favorable convergence rates. Specifically, after $2^{10}$ operation cycles, the \textbf{\texttt{P2LSG}} sequence surpasses the \textbf{\texttt{Niederreiter}} sequence and approaches 
the \textbf{\texttt{Sobol}} sequence in terms of accuracy. For approximate results, the \textbf{\texttt{Sobol}}, \textbf{\texttt{Niederreiter}}, and \textbf{\texttt{P2LSG}} sequences emerge as the top performers.
As can be seen in Fig.~\ref{benchmark_and_mux}~(a), 
the convergence behavior 
of the \textbf{\texttt{\Design}} sequence outperforms other sequences as the length of the bit-stream increases. 

\subsection{Benchmark-II: SC Addition}
\label{benchmark_2}
Next, we 
evaluate the accuracy of the SC \textit{Scaled-Addition}. 
We utilize a 2-to-1 \texttt{MUX} with two 8-bit precision input operands similar to the 
multiplication operation. 
For this SC operation, the two addends (the main inputs of the \texttt{MUX}) are 
correlated ({\small $SCC=1$}), 
while the \texttt{MUX} select input is 
uncorrelated to the addends~\cite{Alaghi_SCC1}. To meet this requirement, we use a random sequence 
to generate the main 
input bit-streams and another sequence to generate the bit-stream corresponding to 
the \texttt{MUX} select input. For two-input addition, a bit-stream corresponding to 0.5 value is generated for the select input. Table~\ref{Table3_add} and Fig.~\ref{benchmark_and_mux} (b) present the accuracy results in terms of 
MAE for different bit-stream lengths. 
Due to using a select input with 2-bit precision (i.e., 0.5 value), accurate output (0.0\% MAE) can be achieved with a bit-stream length of $2^9$ by using sequences such as \textbf{\texttt{Sobol}}, \textbf{\texttt{Niederreiter}}, \textbf{\texttt{Hammersley}}, and \textbf{\texttt{P2LSG}}. 
As can be seen in Table~\ref{Table3_add}, for bit-stream sizes greater than $2^4$, 
the \textbf{\texttt{P2LSG}} sequence achieves 
the minimum MAE among the other sequences. 
By increasing the bit-stream length ($N$), we can see that 
the MAE tends to 
zero for \textbf{\texttt{Sobol}} and \textbf{\texttt{P2LSG}} sequences.

\begin{figure}[t]
  \centering
  \includegraphics[width=230pt]{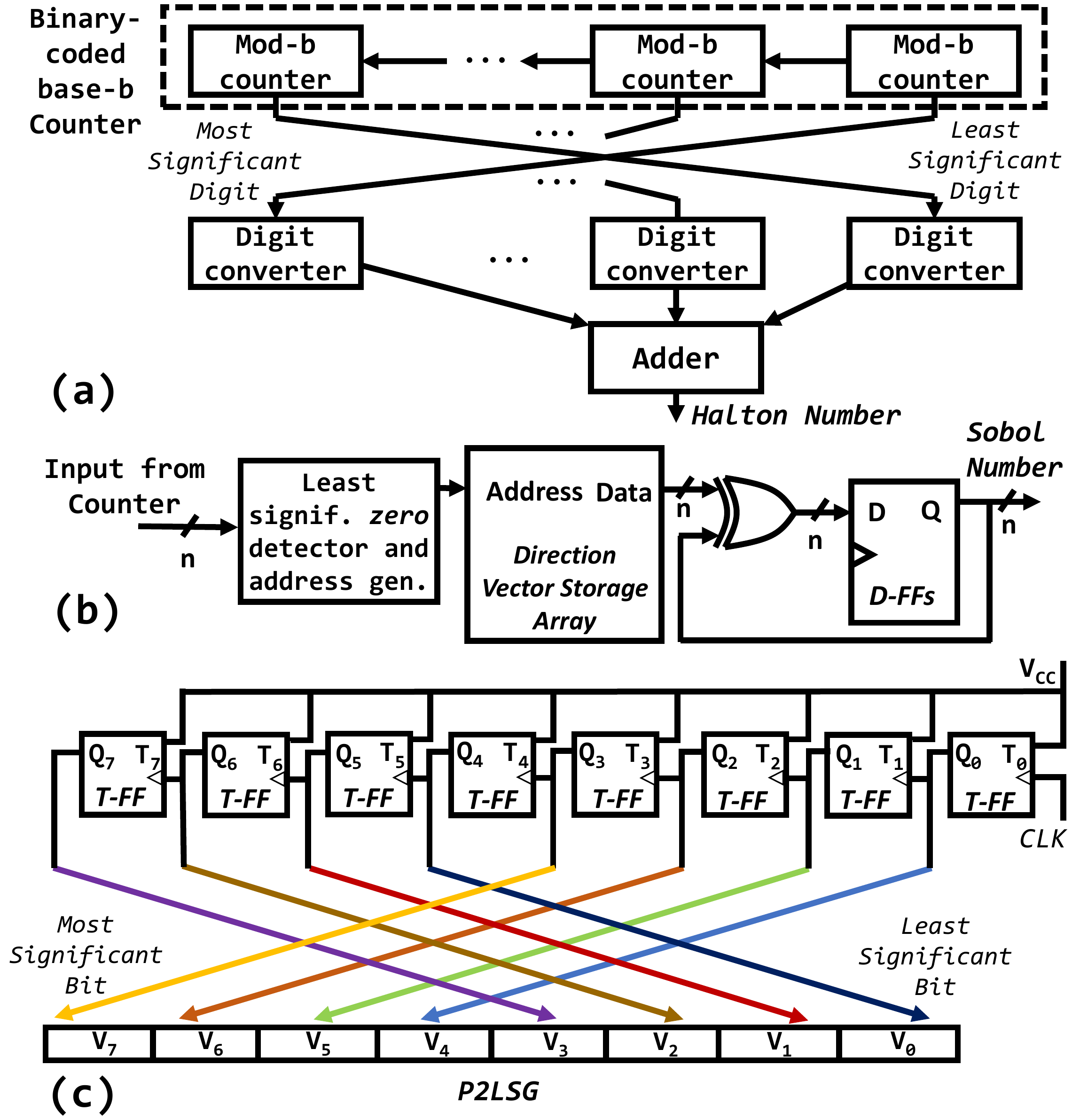}
\vspace{-0.5em}
  \caption{
  (a) The \textbf{\texttt{Halton}} sequence generator implemented in~\cite{6800290}, (b) The \textbf{\texttt{Sobol}} sequence generator proposed in~\cite{Sobol_TVLSI_2018}. (c) Our proposed \textbf{\texttt{\Design}} design for base-16 (\textbf{\texttt{\Design -16}}) as an example; compared to SOTA designs of \textbf{\texttt{Halton}} and \textbf{\texttt{Sobol}} in (a) and (b), \textbf{\texttt{\Design}} utilizes only T-FFs and simple hard-wiring.  
  }
  \label{sobol_halton_ref}
\vspace{-0.5em}
\end{figure}

\section{Proposed Sequence Generator}
\label{new_seq_des}

In this section, we propose a novel hardware design for 
\textbf{\texttt{\Design}} and evaluate its implementation cost compared to prior LD sequence generators.
Alaghi and Hayes~\cite{6800290} implemented 
a \textbf{\texttt{Halton}} sequence generator consisting of mod counters, digit converters, and an adder. Fig.~\ref{sobol_halton_ref} (a) shows the \textbf{\texttt{Halton}}  generator of~\cite{6800290}.
Liu and Han~\cite{Sobol_TVLSI_2018} proposed a \textbf{\texttt{Sobol}} sequence generator by using some 
Direction Vectors (DVs). 
The DVs ({\small ${V_x}(x=0, 1, ..., N-1)$}) are generated using some primitive polynomials and stored in a Direction Vector Array (DVA). 
By employing different DVs, 
different 
\textbf{\texttt{Sobol}} sequences can be produced. 
In their design, 
a priority encoder finds 
the least significant zero (LSZ) in the output of a counter at any cycle. Depending on the position of the LSZ, a DV is selected  
from the DVA. A new \textbf{\texttt{Sobol}} number 
is recursively generated by \texttt{XOR}ing the respective DV and the previous \textbf{\texttt{Sobol}} number. 
Fig.~\ref{sobol_halton_ref} (b) shows the design of this \textbf{\texttt{Sobol}} generator.

Prior work suggested a look-up table-based approach for generating \textbf{\texttt{VDC}} sequences
\cite{TLeeDissertation2019} without a custom hardware design for SC bit-stream generation. 
We propose a low-cost design 
for efficient and lightweight generation of the \textbf{\texttt{P2LSG}} sequences.
Our design uses a {\small $log_2(N)$}-bit counter for generating different sequences of \textit{Powers-of-2} bases up to $N$, where $N$ is the length of the bit-stream.

For a fair comparison with previous random generators and to assess the performance for various image processing applications, 
we target bit-streams of up to 256 (sufficient for representing 8-bit grayscale image data). Therefore, we require up to 256 different random numbers from the 
sequence generator to generate each bit-stream. 
The general algorithm to generate a base-\texttt{B} \textbf{\texttt{VDC}} sequence 
consists of five steps:
\begin{itemize}
    \item [{\footnotesize \textbf{\myCircled{1}}}]Generating an integer 
number.
    \item [{\footnotesize \textbf{\myCircled{2}}}]Converting the integer 
number to its base-\texttt{B} representation.
    \item [{\footnotesize \textbf{\myCircled{3}}}]Reversing the base-\texttt{B} representation.
    \item [{\footnotesize \textbf{\myCircled{4}}}]Converting the base-\texttt{B} representation to a binary number.
    \item [{\footnotesize \textbf{\myCircled{5}}}]Scaling the input number within the {\small $[0, 1)$} interval to the corresponding 8-bit binary number in the {\small $[0, 256)$} range to be connected to the 
binary comparator.

\end{itemize}

The complexity of the hardware design 
for this algorithm is closely tied to the chosen base.  
We classify the hardware designs 
into two categories depending on 
the selected base: 
\textit{Class-I}: those \textit{without Powers-of-2} bases, and \textit{Class-II}: those \textit{with Powers-of-2} bases. 

\subsection{Class-I: \textit{Non-Powers-of-2} Base Generators}

To implement
this type of  \textbf{\texttt{VDC}} sequence generators, 
we combine the first two steps, {\footnotesize \textbf{\myCircled{1}}}and {\footnotesize \textbf{\myCircled{2}}}, by utilizing a base-\texttt{B} counter to generate the integer 
numbers in the corresponding base. 
For instance, a \textit{Binary Coded Decimal} (BCD) counter can be employed for a base-10 representation. Step {\footnotesize \textbf{\myCircled{3}}}is
achieved by hard-wiring. 
Step {\footnotesize \textbf{\myCircled{4}}}
is implemented 
by employing adders and \texttt{MUX}s. This step is relatively complex and 
takes more 
hardware resources compared to the other steps. Step~{\footnotesize \textbf{\myCircled{5}}}can simply be achieved by 
shift operations.

The \textbf{\texttt{Hammersley}} and \textbf{\texttt{Halton}} sequences extend the \textbf{\texttt{VDC}} sequence to higher dimensions, representing each dimension in a different prime base-\texttt{B}. Consequently, the hardware implementation of these sequences falls under this particular type of  sequence generator. 
The need for counters with prime radices and 
base conversion make the \textbf{\texttt{Halton}} sequence generator of~\cite{6800290} complex to implement in hardware.
The hardware limitations of the design of~\cite{6800290} 
motivate us to explore the second class of 
generators for the \textit{Powers-of-2} bases that build 
\textbf{\texttt{\Design}}. 

\subsection{Class-II: \textit{Powers-of-2} Base Generators}

\subsubsection{Sequential Design}

To implement \textit{Powers-of-2} base generators, a binary counter with sufficient bits 
is utilized to represent the desired range of 
integer numbers 
in step {\footnotesize \textbf{\myCircled{1}}}. To convert the value of a binary counter to its base-\texttt{B} representation (step {\footnotesize \textbf{\myCircled{2}}}), we consider groups of {\small $log_2(B)$} bits, starting from the least significant bit. 
If the last group lacks enough bits, some additional `\MVZero'~bits  
are appended via zero padding to ensure it forms a complete group. The reversing operation in step {\footnotesize \textbf{\myCircled{3}}}is done by hard-wiring 
each group of bits, treating them as a single digit in base-\texttt{B}. 
The process of converting a base-\texttt{B} number to its equivalent binary representation is the inverse of step {\footnotesize \textbf{\myCircled{2}}}. In this process, each group or base-\texttt{B} digit is considered as equivalent {\small $log_2(B)$} bits of binary representation and any exceeding bits beyond the counter in step {\footnotesize \textbf{\myCircled{1}}}is discarded. 
Fig.~\ref{sobol_halton_ref}~(c) demonstrates a simple 8-bit precision \textbf{\texttt{P2LSG}} for base-16 (\textbf{\texttt{P2LSG-16}}).

\begin{figure}[t]
  \centering
  \includegraphics[width=\linewidth]{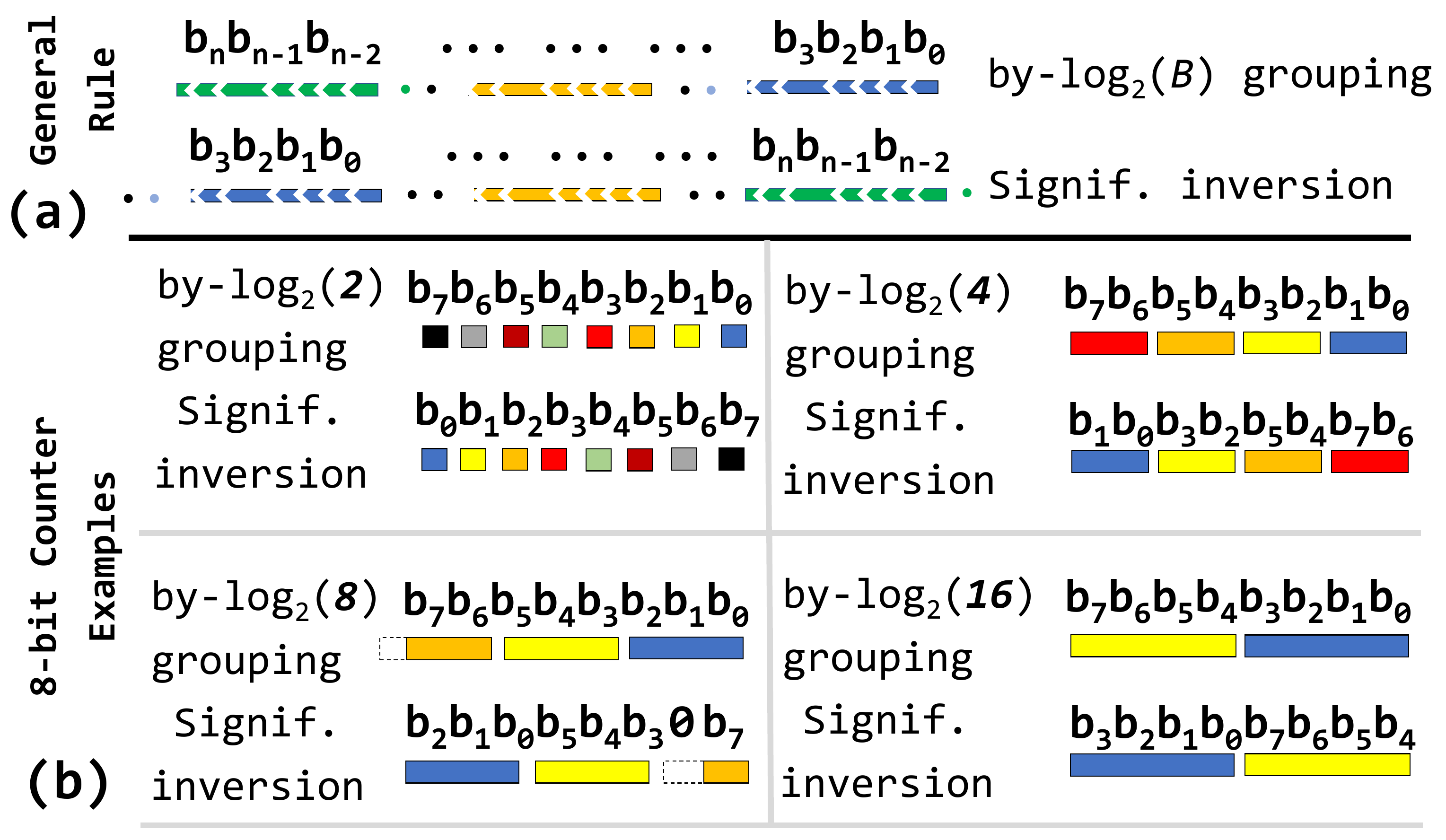}
\vspace{-1.5em}
  \caption{Proposed \textbf{\texttt{P2LSG}}. (a) The general rule to hard-wiring bits for reversing operation (aka., significance inversion), and (b) Example for the 8-bit counter to generate \textbf{\texttt{\Design -2}}, \textbf{\texttt{\Design -4}}, \textbf{\texttt{\Design -8}}, and \textbf{\texttt{\Design -16}} sequences (up to \textbf{\texttt{\Design -256}} is possible).}
  \label{vdcbasep2el}
\end{figure}

An \texttt{Up-Counter} counts up to 255, and the 
target sequence is obtained by significance inversion of each group; the least significant group becomes the most significant group, and vice versa. For this base-16 example, the output {\small $Q_3$} of the {\small $4^{th}$} \texttt{T Flip-Flop} (T-FF) from the right side becomes the most-significant bit. Fig.~\ref{vdcbasep2el}~(a) shows the overall idea behind the proposed \textbf{\texttt{P2LSG}}.
After grouping each bit from the counter, the inversion (via hard-wiring) reverses the bit significance; the new binary output is ready for comparison in the SNG block.  Fig.~\ref{vdcbasep2el}~(b) illustrates examples of different bases.

\begin{figure}[t]
  \centering
  \includegraphics[width=\linewidth]{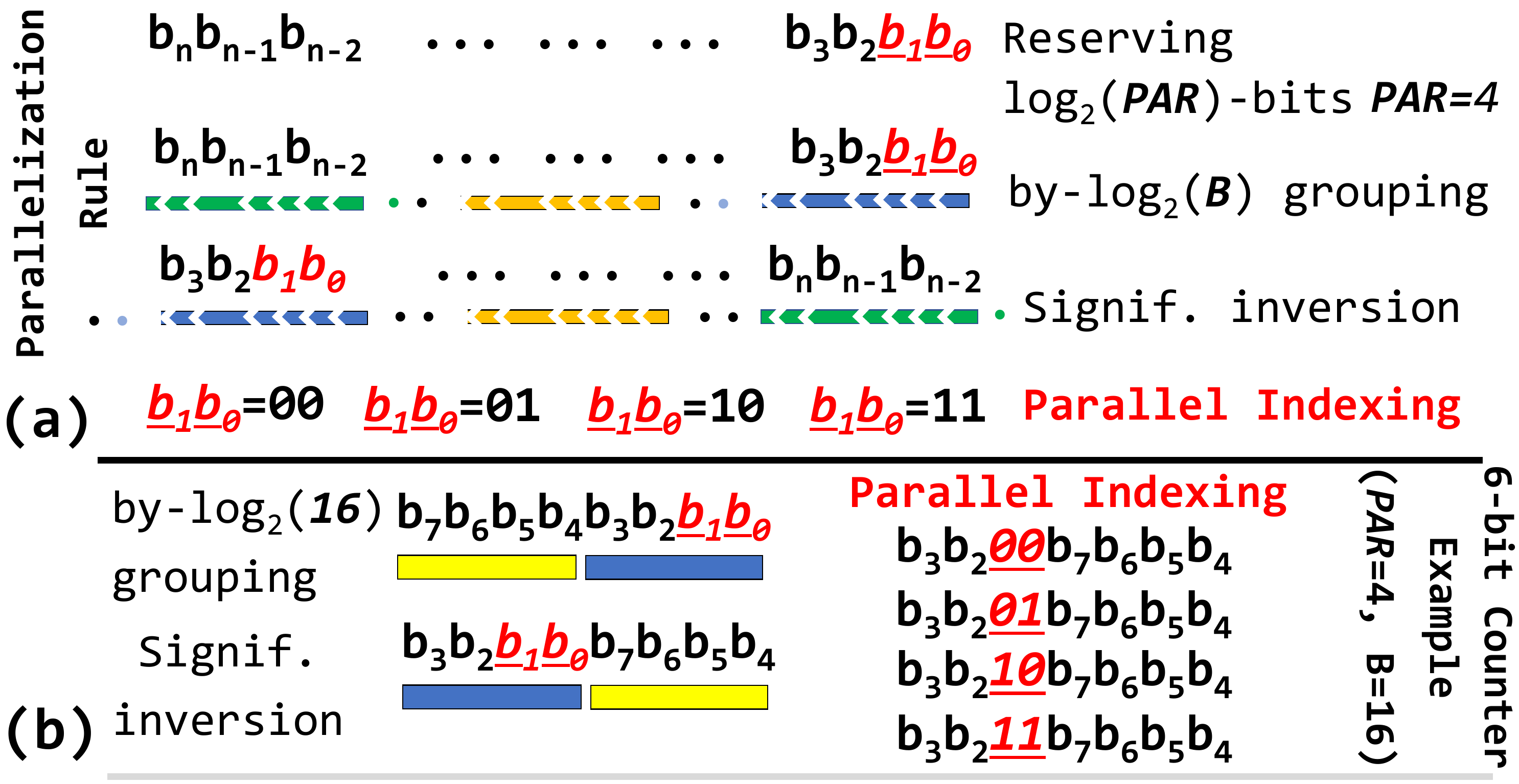}
\vspace{-1.5em}
  \caption{Parallel \textbf{\texttt{\Design}} sequence generator. (a) The general rule to assign parallel indexing bits, and (b) \textbf{\texttt{\Design -16}} example with {\scriptsize $PAR=4$} concurrent sequence generation.}
  \label{VDC_Design2}
\vspace{-1.em}
\end{figure}

\begin{table}[b]
\centering
\vspace{-1.em}
\caption{Hardware Cost Comparison for Generating Two Different\\ 8-bit Random Sequences.}
\vspace{-0.8em}
\setlength{\tabcolsep}{3pt}
\begin{tabular}{|c|c|c|c|} 
\hline
\textbf{ SNG } & \begin{tabular}[c]{@{}c@{}}\textbf{ Area}\\ ({\boldmath $\mu$}\textbf{m\textsuperscript{2}})\end{tabular} & \begin{tabular}[c]{@{}c@{}}\textbf{ Power}\\({\boldmath $\mu$}\textbf{W})\end{tabular} & \begin{tabular}[c]{@{}c@{}}\textbf{ CPL}\\(\textbf{ns})\end{tabular} \\ 
\hline
\textbf{\texttt{Sobol\#2 \& Sobol\#3}} \cite{Najafi_TVLSI_2019} & $2\times781$ & $2\times45.15$  & 0.68 \\ 
\hline
\textbf{\texttt{Halton\#1 \& Halton\#2}} & $130+450$ & $15.15+35.30$  & 1.06 \\ 
\hline
\textbf{\texttt{P2LSG-4 \& P2LSG-16}} & \textbf{163} & \textbf{16.05}  & \textbf{0.49} \\
\hline
\end{tabular}
\label{TableHW}
\vspace{-0.5em}
\justify{\scriptsize{
 \textbf{\texttt{Halton\#1}} and \textbf{\texttt{Halton\#2}} are generated using \textbf{\texttt{VDC-2}} and \textbf{\texttt{VDC-3}}, respectively. In this case, the base-$2$ and base-$3$ counters are utilized to generate the \textbf{\texttt{Halton}} sequence. It is worth noting that designing counters with prime bases (except base-$2$) can lead to a significant hardware cost. As for \textbf{\texttt{P2LSG-4}} and \textbf{\texttt{P2LSG-16}}, they employ \textbf{\texttt{VDC-4}} and \textbf{\texttt{VDC-16}}, respectively. Any \textit{Powers-of-2} bases can be utilized to generate \textbf{\texttt{\Design}} sequences with the similar design in Fig.~\ref{sobol_halton_ref} except with different hard-wiring schemes. (CPL: Critical Path Latency)
}}

\end{table}

\subsubsection{Parallel Design}

Our proposed \textit{Class-II} \textbf{\texttt{P2LSG}} can also operate in parallel. 
Fig.~\ref{VDC_Design2} illustrates how more than one number of 
a \textbf{\texttt{P2LSG}} sequence (in any base) can be generated in parallel at any cycle. 
Let us define {\small $PAR$} as the number of sequence elements to be generated in parallel. 
First, {\small $log_2(PAR)$} bits are reserved at the least significant positions. The remaining bits require a reduced precision counter (e.g., {\small $8$$\rightarrow$$6$} in Fig.~\ref{VDC_Design2}). 
At any clock cycle, the reserved bits are filled
with 
{\small $2^{(log_2(PAR))}$} possible logic values (parallel indexing). Fig.~\ref{VDC_Design2}~(a) shows an example for {\small $PAR=4$}. In this example, each output repeats 
four times to fill the reserved bits with \MVZero\MVZero, \MVZero\MVOne, \MVOne\MVZero, and \MVOne\MVOne. 
The outputs at any cycle 
produce four consecutive 
numbers. 
Fig.~\ref{VDC_Design2}~(b) illustrates another example of {\small $PAR=4$} for 
\textbf{\texttt{\Design -16}}.

Table~\ref{TableHW} compares the hardware cost of generating 
\textbf{\texttt{Sobol}} and \textbf{\texttt{Halton}}
sequences with the proposed sequence generator. 
We report the hardware area, power consumption, and critical path latency (CPL) for each case. 
We synthesized the designs 
using the Synopsys Design Compiler v2018.06 with the 45nm FreePDK gate library~\cite{FreePDK45}. 
The reported numbers in 
Table~\ref{TableHW} demonstrate that the proposed sequence generator surpasses the \textbf{\texttt{Sobol}} 
and \textbf{\texttt{Halton}} sequence generators in terms of hardware efficiency.

\section{SC Image and  Video Case Studies}
\label{sc_image}

In this section, we evaluate the performance 
and the hardware efficiency of the proposed \textbf{\texttt{P2LSG}} in two SC image and video processing case studies. 
Prior work has used SC for low-cost implementation of different computer vision tasks from depth perception to interpolation~\cite{Weikang_2011, accurateYet, 7437471, 7858369, SC_Image_2014, 7960479}. 
We first evaluate the proposed sequence generator in an interpolation and image scaling application and then implement and study its effectiveness in a novel SC circuit for scene merging video processing, which we propose 
for the first time in the literature. 

\subsection{Interpolation and Image Scaling}
Interpolation refers to the process of estimating or calculating values between two known data points. Linear interpolation 
is a method used to estimate values between two known values based on a linear relationship \cite{1288196}. It assumes a straight line between the available values and calculates intermediate values along that line. In 
image processing, linear interpolation is used to estimate pixel values between two neighbouring pixels. It is commonly employed when performing operations such as rotation, translation, or affine transformations on images. Bilinear interpolation is a specific case of linear interpolation applied in two dimensions. Instead of estimating values along a straight line, it estimates values within a two-dimensional grid of pixels \cite{1409828}. Bilinear interpolation considers the four nearest pixels to the target location and calculates a weighted average based on their values. The weights are determined by the distances between the target location and the surrounding pixels; thereby, an image scaling task can be performed~\cite{press1995numeric}.

\begin{figure*}[t]
  \centering
  \includegraphics[width=\linewidth]{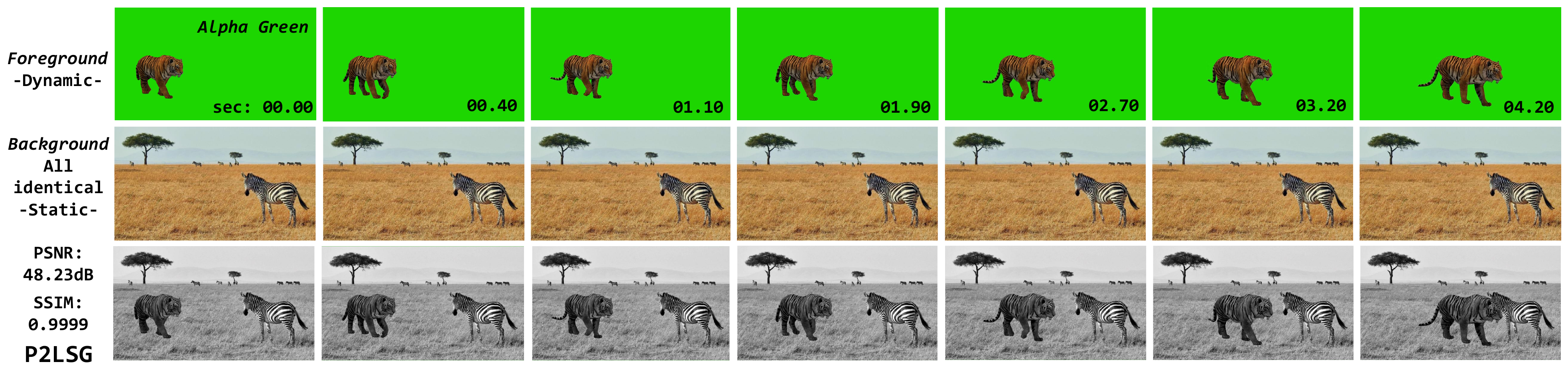}
\vspace{-1.5em}
  \caption{SC-based video processing for scene merging. The foreground video with green background is embedded into a static jungle background  picture. The figure shows several frames from the first 04.20 sec of the video in the test. For $N$=256, the results of \textbf{\texttt{P2LSG}}-based and \textbf{\texttt{Sobol}}-based SC designs alongside the reference binary-based conventional processing can be accessed from: \blue{\url{https://github.com/mehranUL/LD_Seqs_SC/blob/main/Processed_videos/}}. 
  }
  \label{Video_figure}
\end{figure*}

Assume we have an original image, $I$, with pixel values represented by a 2-D array. We want to estimate the pixel value at a non-integer coordinate {\small $(\mathrm{x},\mathrm{y})$} in the image. The four surrounding pixels to consider are {\small $(x_1, y_1)$}, {\small $(x_1, y_2)$}, {\small $(x_2, y_1)$}, and {\small $(x_2, y_2)$}, where {\small $(x_1, y_1)$} represents the pixel at the bottom-left corner of the target location, and {\small $(x_2, y_2)$} represents the pixel at the top-right corner. Let us denote the pixel values as {\small $I(\mathrm{x},\mathrm{y})$}, {\small $I(x_1, y_1)$}, {\small $I(x_1, y_2)$}, {\small $I(x_2, y_1)$}, and {\small $I(x_2, y_2)$}. The bilinear interpolation formula to estimate the pixel value {\small $I(\mathrm{x},\mathrm{y})$} is as follows: {\small $I(\mathrm{x},\mathrm{y}) = (1 - u)(1 - v) \times I(x_1, y_1) + (1 - u)v \times I(x_1, y_2) + u(1 - v) \times I(x_2, y_1) + uv \times I(x_2, y_2)$}, where {\small $u = x - x_1$} (fractional distance between {\small $x$} and {\small $x_1$}) and {\small $v = y - y_1$} (fractional distance between {\small $y$} and {\small $y_1$}). The values {\small $(1 - u)(1 - v)$}, {\small $(1 - u)v$}, {\small $u(1 - v)$}, and {\small $uv$} are the weights assigned to each surrounding pixel. These weights represent the contribution of each pixel to the interpolated value. The interpolation formula can be compared to a multiplication-based SC \texttt{MUX} structure~\cite{10.1145/3491213}, where neighbouring pixels are fed into the main \texttt{MUX} inputs, and the location information is fed into the selection ports. In this scenario, 
a 4-to-1 \texttt{MUX} 
can be expressed in terms of probabilities as follows: {\small $P_{\boldsymbol{I}(\mathrm{x},\mathrm{y})} = (1-P_{\boldsymbol{u}})(1-P_{\boldsymbol{v}})P_{\boldsymbol{I}_{11}}$} \linebreak {\small $+ (1-P_{\boldsymbol{u}})(P_{\boldsymbol{v}})P_{\boldsymbol{I}_{12}} + (P_{\boldsymbol{u}})(1-P_{\boldsymbol{v}})P_{\boldsymbol{I}_{21}} + (P_{\boldsymbol{u}})(P_{\boldsymbol{v}})P_{\boldsymbol{I}_{22}}$}.

\begin{figure}[t]
  \centering
  \includegraphics[width=\linewidth]{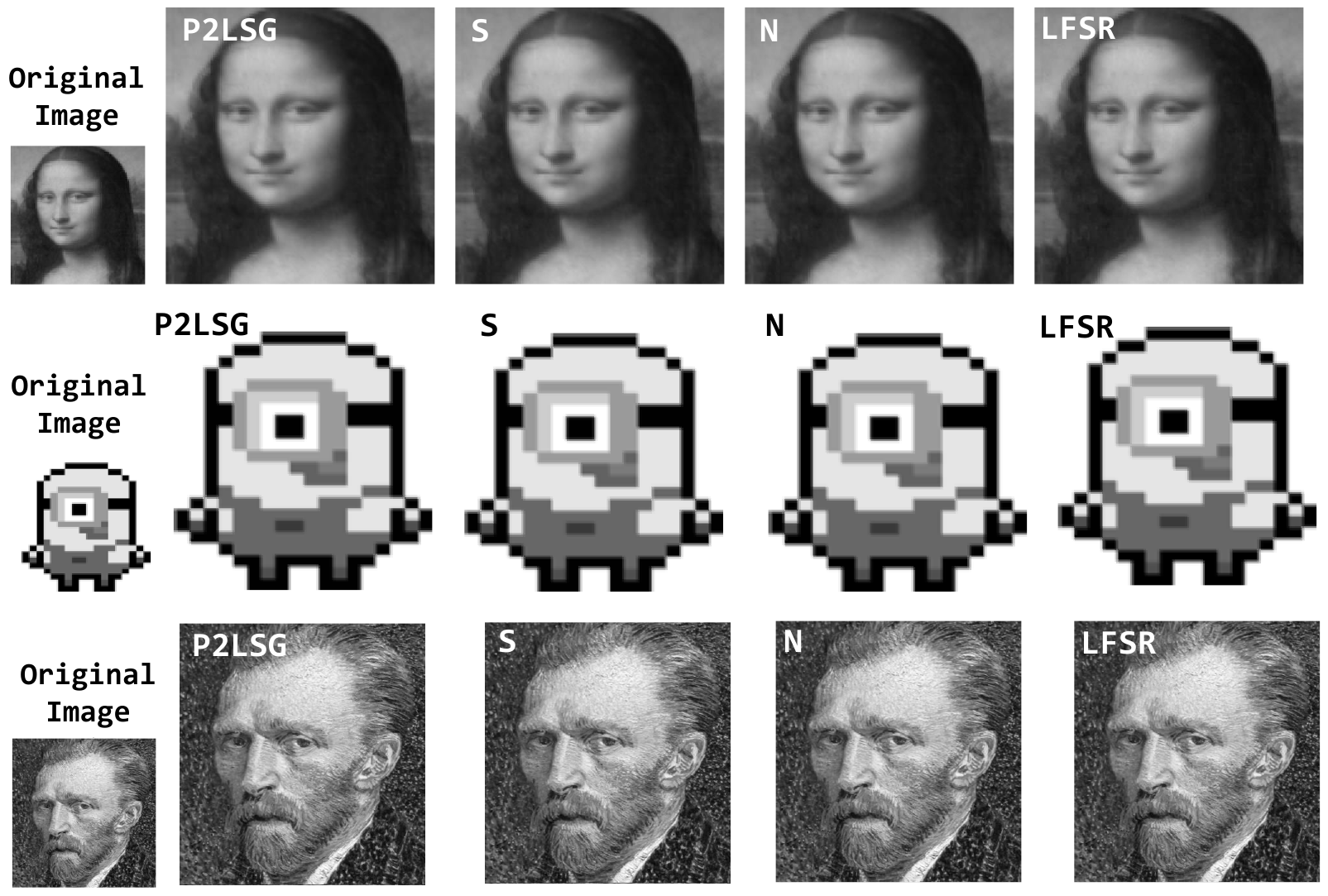}
\vspace{-1.5em}
  \caption{Visual results of SC image scaling using different SNGs (with \textbf{\texttt{P2LSG}}, \textbf{\texttt{Sobol}}, \textbf{\texttt{Niederreiter}}, or LFSR) and a 4-to-1 \texttt{MUX}.}
  \label{mona_lisa_van_gogh}
\vspace{-1em}
\end{figure}

\begin{table}[t]
\centering
\caption{SC Image Scaling with Different Random Sequences. 
}
\vspace{-0.8em}
\setlength{\tabcolsep}{3pt}
\begin{tabular}{|c|c|c|c|} 
\hline
\multicolumn{4}{|c|}{\textbf{Image Scaling (}\textit{Scale Factor}=2\textbf{)}} \\ 
\hline
\textbf{\textbf{\texttt{P2LSG}}} & \textbf{\texttt{Sobol (S)}} & \textbf{\texttt{Niederreiter(N)}} & \textbf{LFSR} \\ 
\hline
\multicolumn{4}{|c|}{\textbf{\textit{Mona Lisa}}} \\ 
\hline
\begin{tabular}[c]{@{}c@{}}PSNR: \textbf{46.62dB}  \\SSIM: \textbf{0.9958}\end{tabular} & \begin{tabular}[c]{@{}c@{}}PSNR: 46.25dB  \\SSIM: 0.9948\end{tabular} & \begin{tabular}[c]{@{}c@{}}PSNR: 46.37dB \\SSIM: 0.9949\end{tabular} & \begin{tabular}[c]{@{}c@{}}PSNR: 44.63dB  \\SSIM: 0.9892\end{tabular} \\ 
\hline
\multicolumn{4}{|c|}{\textbf{\textit{Minion}}} \\ 
\hline
\begin{tabular}[c]{@{}c@{}}PSNR: \textbf{39.36dB}  \\SSIM: \textbf{0.9975}\end{tabular} & \begin{tabular}[c]{@{}c@{}}PSNR: 39.26dB  \\SSIM: 0.9926\end{tabular} & \begin{tabular}[c]{@{}c@{}}PSNR: 39.30dB \\SSIM: 0.9958\end{tabular} & \begin{tabular}[c]{@{}c@{}}PSNR: 38.50dB  \\SSIM: 0.9964\end{tabular} \\ 
\hline
\multicolumn{4}{|c|}{\textbf{\textit{Van Gogh}}} \\ 
\hline
\begin{tabular}[c]{@{}c@{}}PSNR: \textbf{43.10dB} \\SSIM: \textbf{0.9958}\end{tabular} & \begin{tabular}[c]{@{}c@{}}PSNR: 42.91dB \\SSIM: 0.9921\end{tabular} & \begin{tabular}[c]{@{}c@{}}PSNR: 42.99dB \\SSIM: 0.9923\end{tabular} & \begin{tabular}[c]{@{}c@{}}PSNR: 41.65dB \\SSIM: 0.9880\end{tabular} \\
\hline
\end{tabular}
\label{Tab:ImageScalingResults}
\end{table}

Fig.~\ref{mona_lisa_van_gogh} visually demonstrates the outputs of 2$\times$ image scaling with an SC circuit 
composed of SNGs for data conversion 
and a 4-to-1 \texttt{MUX} unit. 
Table~\ref{Tab:ImageScalingResults} presents the performance results in terms of peak-signal-to-noise ratio (PSNR) and structural similarity (SSIM). 
We evaluated the SC circuit for the cases of using 
the \textbf{\texttt{P2LSG}}, \textbf{\texttt{Sobol (S)}}, \textbf{\texttt{Niederreiter (N)}}, and LFSR random sequences in the SNG units. 
We processed the \textit{Mona Lisa}, \textit{Minion}, and \textit{Van Gogh} images as the test images. The results shown in Fig.~\ref{mona_lisa_van_gogh} and Table~\ref{Tab:ImageScalingResults} show the superior performance of the \textbf{\texttt{P2LSG}} sequences. 
We further evaluated the performance and energy consumption in 45nm CMOS technology 
when processing the \textit{Mona Lisa} image (107 $\times$ 104 image size) for the two cases of \textbf{\texttt{P2LSG}} and \textbf{\texttt{Sobol}}. 
Table~\ref{final_hw_table} reports the results. 
As reported, the non-parallel and the 4$\times$ parallel designs of the \textbf{\texttt{P2LSG}}-based implementation save area by 64\% and 55\%, energy by 67\% and 85\%, and CPL by 13\% and 22\% compared to the non-parallel and 4$\times$ parallel \textbf{\texttt{Sobol}}-based implementation, respectively. 

\begin{table}[t]
\centering
\caption{Hardware Cost Comparison 
of the Image and\\ Video Processing Case Studies. }
\vspace{-0.8em}
\setlength{\tabcolsep}{3.5pt}
\begin{tabular}{|c|c|c|c|c|c|} 
\hline
\multirow{2}{*}{$N$=256} & \begin{tabular}[c]{@{}c@{}}\textbf{Area} \\({\boldmath $\mu$}\textbf{m\textsuperscript{2}})\end{tabular} & \begin{tabular}[c]{@{}c@{}}\textbf{*Energy} \\(\textbf{pJ})\end{tabular} & \begin{tabular}[c]{@{}c@{}}\textbf{*CPL} \\~(\textbf{ns})\end{tabular} & \begin{tabular}[c]{@{}c@{}}\textbf{Total Energy} \\({\boldmath{$\mu$}}\textbf{J})\end{tabular} & \begin{tabular}[c]{@{}c@{}}\textbf{Runtime} \\\end{tabular} \\ 
\cline{2-6}
 & \multicolumn{5}{c|}{\textbf{Image Scaling}} \\ 
\hline
\textbf{\texttt{Sobol}} \cite{Najafi_TVLSI_2019} & 2017 & 17.55 & 366 & 0.781 & 16.3 ms \\ 
\hline
Parallel 4$\times$ \textbf{\texttt{Sobol}} & 4548 & 16 & 91.5 & 0.713 & 4.07 ms \\ 
\hline
\textbf{\texttt{P2LSG}} & \textbf{715} & \textbf{5.6} & \textbf{317} & \textbf{0.25} & \textbf{14.12 ms}  \\ 
\hline
Parallel 4$\times$ \textbf{\texttt{P2LSG}} & \textbf{2040} & \textbf{2.4} & \textbf{71} & \textbf{0.107} & \textbf{3.16 ms}  \\ 
\hline
~ & \multicolumn{5}{c|}{\textbf{Scene Merging}} \\ 
\hline
\textbf{\texttt{Sobol}} \cite{Najafi_TVLSI_2019}  & 1787 & 16.6 & 353 & 1568 & 33.3 s \\ 
\hline
Parallel 4$\times$ \textbf{\texttt{Sobol}} & 3628 & 15.1 & 88 & 1424 & 8.33 s \\ 
\hline
\textbf{\texttt{P2LSG}} & \textbf{485} & \textbf{4.7} & \textbf{304} & \textbf{443} & \textbf{28.7 s}  \\ 
\hline
Parallel 4$\times$ \textbf{\texttt{P2LSG}} & \textbf{1120} & \textbf{1.48} & \textbf{68} & \textbf{140} & \textbf{6.4 s}  \\
\hline
\end{tabular}
\label{final_hw_table}
\vspace{-0.6em}
\justify{\scriptsize{
*Energy and *CPL are for producing each output pixel. Bit-stream Length ($N$) is 256. (CPL: Critical Path Latency)
}}

\end{table}

\subsection{SC Video Processing: Scene Merging}

As the second case study, 
we introduce an SC circuit for scene merging in video processing. 
The goal 
is to merge two different scenes in a video.
One scene is 
a static image (background), while the other is a moving video sequence (foreground). To composite the images, the background image is processed with a transparent foreground image (green \textit{alpha} channel) using the formula: {\small $Merged \ Pixel = Background \ Pixel \times (1-alpha) + Foreground \ Pixel \times alpha$} \cite{Sun_2021_CVPR}. Here, the \textit{alpha} channel, which depends on the dynamic movement of the moving object in the green background of the video, is updated frame by frame. Considering that the \textit{alpha} value should be within the {\small $[0, 1]$} interval, the formula can be seen as a \texttt{MUX} with inputs being the foreground and background scenes and the selection port being the \textit{alpha} channel: {\small $MUX = P_{\boldsymbol{X_1}}(1-P_{\boldsymbol{S}}) + P_{\boldsymbol{X_2}}P_{\boldsymbol{S}}$}, where {\small $P_{\boldsymbol{X_1}}$} and {\small $P_{\boldsymbol{X_2}}$} are the circuit inputs and {\small $P_{\boldsymbol{S}}$} is 
the select input~\cite{10.1145/3491213}. Based on this analogy, the provided video in Fig.~\ref{Video_figure}, which features a moving tiger, is merged with an African jungle background. Using the \textbf{\texttt{P2LSG}} sequence for data conversion and a 2-to-1 \texttt{MUX} as the SC circuit, the processed video achieves a PSNR of $48.23dB$ and an SSIM of $0.999$. The video has a duration of $6.06$ seconds and was generated at a frame rate of 30 frames per second. 
We also achieved similar
PSNR and SSIM values ($48.13dB$ and $0.9999$) when using the \textbf{\texttt{Sobol}}-based SNG for data conversion. As the hardware results presented in Table~\ref{final_hw_table} demonstrate, the non-parallel and 4$\times$ parallel \textbf{\texttt{P2LSG}} design provide 73\% and 69\% lower area, 72\% and 90\% lower energy consumption, and 14\% and 23\% lower runtime compared to the non-parallel and 4$\times$ parallel \textbf{\texttt{Sobol}}-based design.  

\section{Conclusions}
\label{conclusions}
This study explores new design possibilities for SC by analyzing some well-known 
random sequences in the literature. As a promising random sequence for the SNG unit of SC systems, 
we evaluated the performance of the \textbf{\texttt{P}}owers-of-\textbf{\texttt{2}} \textbf{\texttt{L}}ow-Discrepancy
\textbf{\texttt{S}}equence \textbf{\texttt{G}}enerator (\textbf{\texttt{P2LSG}})
sequences. 
We proposed a lightweight hardware design for \textbf{\texttt{P2LSG}} sequence generation. The proposed generator provides a higher hardware efficiency compared to the SOTA LD sequence generators.  
We evaluated the performance of the \textbf{\texttt{P2LSG}}-based SNGs in an 
image scaling and a scene merging video processing case study. Our performance evaluation and hardware cost comparison show comparable or better numbers compared to the SOTA. 
Our finding opens 
possibilities for incorporating the \textbf{\texttt{P2LSG}} sequences in other emerging 
paradigms that require orthogonal 
vectors, such as hyperdimensional computing. 
The \textbf{\texttt{P2LSG}} sequences can be utilized 
to address the computational needs and improve the performance of such paradigms. We leave studying this aspect for our future work.

\bibliographystyle{IEEEtran}
\bibliography{bibliography,Hassan}

\end{document}